\documentclass[aps,pra,twocolumn,amsmath,amssymb,superscriptaddress,longbibliography]{revtex4-1}
\usepackage[english]{babel}

\usepackage{amssymb}
\usepackage{dcolumn}
\usepackage{bm}
\usepackage{graphicx}
\usepackage{amsmath}

\usepackage{graphicx}        
\graphicspath{{pict/}{}}

\usepackage{dcolumn}
\usepackage{bm}
\usepackage[dvipdfm, pdfstartview=FitH, CJKbookmarks=true, bookmarksnumbered=true, bookmarksopen=true, colorlinks=true, pdfborder=001, citecolor=blue, linkcolor=blue, urlcolor=blue, linktocpage=true] {hyperref}

\begin{document}

\title{Floquet control of global $PT$ symmetry in 2D arrays of quadrimer waveguides}

%
%
%

\author{Bo Zhu}
\affiliation{Institute of Mathematics and Physics, Central South  University of Forestry and Technology, Changsha 410004, China}
\affiliation{School of Physics and Astronomy, Sun Yat-Sen University (Zhuhai Campus), Zhuhai 519082, China}

\author{Honghua Zhong}
\altaffiliation{hhzhong115@163.com}
\affiliation{Institute of Mathematics and Physics, Central South  University of Forestry and Technology, Changsha 410004, China}
\affiliation{Graduate School, China Academy of Engineering Physics, Beijing 100193, China}

\author{Jun Jia}
\affiliation{Department of Physics, Jishou University, Jishou 416000, China}

\author{Fuqiu Ye}
\affiliation{Department of Physics, Jishou University, Jishou 416000, China}

\author{Libin Fu}
\altaffiliation{lbfu@gscaep.ac.cn}
\affiliation{Graduate School, China Academy of Engineering Physics, Beijing 100193, China}


\date{\today}

\begin{abstract}
Manipulating the global $PT$ symmetry of a non-Hermitian composite system is a rather significative and challenging task.
Here, we investigate Floquet control of global $PT$ symmetry in 2D arrays of quadrimer waveguides with transverse periodic structure along $x$-axis and longitudinal periodic modulation along $z$-axis.
For unmodulated case with inhomogeneous inter- and intra- quadrimer coupling strength $\kappa_1\neq\kappa$, in addition to conventional global $PT$-symmetric phase and $PT$-symmetry-breaking phase, we find
that there is exotic phase where global $PT$ symmetry is broken under open boundary condition, whereas it still is unbroken under periodical boundary condition.
The boundary of phase is analytically given as $\kappa_1\geq\kappa+\sqrt{2}$ and $1\leq\gamma\leq2$, where there exists a pair of zero-energy edge states with purely imaginary energy
eigenvalues localized at the left boundary, whereas other $4N-2$ eigenvalues are real. Especially,
the domain of the exotic phase can be manipulated narrow and even disappeared by tuning modulation parameter. More interestingly,
whether or not the array has initial global $PT$ symmetry, periodic modulation not only can restore the broken global $PT$ symmetry, but also can control it by tuning modulation amplitude. Therefore, the global property of transverse periodic structure of such a 2D array can be manipulated
by only tuning modulation amplitude of longitudinal periodic modulation.
\end{abstract}

\maketitle

\section{INTRODUCTION}
Global parity-time ($PT$) symmetry
plays a key role in determining the real energy spectrum, topological character and transport property of non-Hermitian composite system  \cite{jyang, rott17,Moiseyev11, konotop,bender,bender2,mahaux,morfon17}.
There are many interesting physical phenomena related to global $PT$ symmetry, including topological bound state \cite{weiman2017}, edge-mode lasing \cite{parto18}, anomalous edge states \cite{rivo17} and anisotropic transmission
resonances \cite{ge2012}.
A characteristic property of global $PT$-symmetric system is the existence of a phase transition (spontaneous global $PT$-symmetry-breaking) from the unbroken to broken-$PT$-symmetric phase whenever the gain/loss parameter exceeds a certain threshold. This has been experimentally demonstrated in synthetic photonic lattices \cite{bersch,weiman2017} and cavity laser arrays \cite{gao}.
Therefore, an important issue in a global $PT$-symmetric system is the ability to control and tune this phase transition.

Optical structures constructed
by arrays of coupled dimers \cite{ nodal2018,longhi2013,bender2014,longhi16,
zezy12,dmitr10,yang18,ramez12,bara13}, trimer \cite{ramez2010} or quadrimers\cite{kalo2016, konotop12,alex17}, provide a fertile ground to observe and utilize notions of global $PT$ symmetry. Global $PT$ symmetry of such  system will
require precise relation between various on-site energies and
coupling symmetry between the building blocks, and hence
becomes extremely fragile in the presence of disorder, impurities and boundaries which can support localized modes \cite{zheng2010,bendix10,scott10,li10,bendix09,nguyen16}.
One of the most significant features of such optical structure is the boundary condition
dependence,
where system under periodic boundary condition (PBC) and open boundary condition (OBC) have dramatically different energy spectra, and the zero-energy edge states (topologically nontrivial phase) related to spontaneous global $PT$-symmetry-breaking transition can appear. These have received great research interests in a class of photonic arrays of
$PT$-symmetric dimers described by the
non-Hermitian
Su-Schrieffer-Heeger (SSH) model \cite{zhu14,klett17,lieu18,hou19, yuc18}.
It was shown that there is no universal correlation between spontaneously global $PT$-symmetry-breaking and the topologically nontrivial phase, and only the symmetry of the individual edge states can decide whether their presence has an influence on the global $PT$ symmetry \cite{klett17}.
Thus the key outstanding question is: What is the accurate parametric region of the boundary influencing global $PT$ symmetry?

Recently, based
on the high-frequency Floquet method to rescale the coupling
strength, periodic modulations have been proposed to control
$PT$ symmetry of single optical dimer \cite{Moiseyev,lian,jog,gong, lee,zhouz,wu2017,song2019}. It has been
found that manipulation of $PT$-phase transition can be achieved
by adjusting
modulation parameter. The Floquet $PT$-symmetric system also has been realized for two coupled LC resonators with balanced gain and loss \cite{mahboo}. However, these previous works mainly consider that
out-of-phase periodic modulations were introduced on
complex on-site energies or intra-dimer coupling strength. This is a rather challenging task in optical experiment, as
the balanced gain and loss and periodic modulation in the complex refractive index must
be tuned simultaneously. Therefore, the protocol that the additional modulated dimer is introduced to constitute a periodically modulated $PT$-symmetric quadrimer may be more easily operate in experiment.
Especially, the study on
Floquet control of global $PT$ symmetry in such a composite array
is still lacking.

In this work, we address above these important questions by investigating global $PT$ symmetry and its Floquet control in 2D arrays of quadrimer waveguides with transverse periodic structure along $x$-axis and longitudinal periodic modulation along $z$-axis, whose single quadrimer is coupled by a $PT$-symmetric dimer and a periodically modulated dimer.
Our main results are:

i) There exists an exotic phase where global $PT$ symmetry is broken under OBC, whereas it still is unbroken under PBC for unmodulated case with asymmetric coupling between inter- and intra- quadrimer.
The boundary of phase is analytically given as $\kappa_1\geq\kappa+\sqrt{2}$ and $1\leq\gamma\leq2$, where there exists a pair of zero-energy edge states with purely imaginary energy
eigenvalues localized at the left boundary, whereas other $4N-2$ eigenvalues are real.

ii) The system may support the triple point, where three phases of robust global-$PT$-symmetry (R$PT$), boundary influencing global-$PT$-symmetry (BI$PT$), and broken global-$PT$-symmetry (B$PT$) touch. Thus depending on how parameters change in vicinity of the triple point, breaking of global $PT$ symmetry can occur in two different ways: R$PT$$\rightarrow$ B$PT$, or R$PT$$\rightarrow$BI$PT$$\rightarrow$B$PT$.

iii) The domain of the exotic phase can be manipulated narrow and even disappeared by tuning modulation parameter $A/\omega$. More interestingly,
whether or not the array has initial global $PT$ symmetry, periodic modulation not only can restore the broken global $PT$ symmetry, but also can control it by tuning modulation amplitude.

iv) The global property of transverse periodic structure of such a 2D array can be manipulated
by only tuning modulation amplitude of longitudinal periodic modulation, which provides a promising approach for designing and manipulating optical material and may has specific technological importance.

\begin{figure}[htb]
\begin{center}
\includegraphics[bb= 142 116 766 369, clip, scale=0.4]{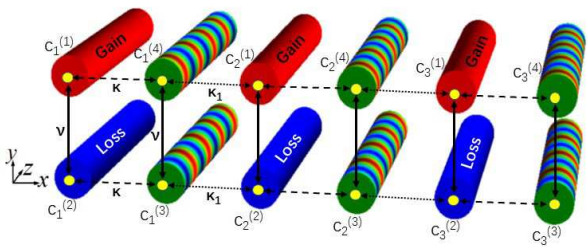}
\caption{(Color online) Schematic diagram of a 2D waveguide array comprised by periodically modulated $PT$-symmetric quadrimers. The periodic change of color along the $z$-axis denotes the periodic modulation $A sin(\omega z)$.
The intra-dimer coupling strength $\nu$ is tuned by adjusting the
center-to-center waveguide spacing along $y$-axis $\Delta y$; the intra-quadrimer coupling strength $\kappa$
and the inter-quadrimer coupling strength $\kappa_1$ are tuned by intermittently adjusting the
center-to-center waveguide spacing along $x$-axis $\Delta x$.} \label{fig0}
\end{center}
\end{figure}

\section{The model}
We consider a 2D array comprised by $N$ quadrimers waveguides, whose single quadrimer is coupled by a $PT$-metric dimer and a periodically modulated dimer, see Fig.\ref{fig0}.
Within a tight-binding model with nearest-neighbor couplings, light dynamics in the
optical structure along the propagation axis $z$ are described by the following coupled-mode equation
\begin{eqnarray} \label{dyequation}
i\frac{\partial \psi(z)}{\partial z}= H_N(z) \psi(z),
\end{eqnarray}
with the Hamiltonian
\begin{eqnarray} \label{model}
H_{N}(z)=\left(\begin{array}{ccccccc}
h_1&\sigma_+&0&\cdots&0&0&0 \\
\sigma_-&h_2&\sigma_+&0&\cdots&0&0 \\
0&\sigma_-&h_3&\sigma_+&0&\cdots&0 \\
\vdots&\ddots&\ddots&\ddots&\ddots&\ddots&\vdots \\
0&\cdots&0&\sigma_-&h_{N-2}&\sigma_+&0 \\
0&0&\cdots&0&\sigma_-&h_{N-1}&\sigma_+ \\
0&0&0&\cdots&0&\sigma_-&h_{N} \\
\end{array} \right),
\end{eqnarray}
where
\begin{eqnarray}\label{qusidm}
h_n=\left(\begin{array}{cccc}
i \gamma&\nu&0&\kappa \\
\nu&-i \gamma&\kappa&0 \\
0&\kappa&-A sin(\omega z)&\nu \\
\kappa&0&\nu&A sin(\omega z) \\
\end{array} \right),
\end{eqnarray}
\begin{eqnarray}
\sigma_-=\left(\begin{array}{cccc}
0&0&0&\kappa_1 \\
0&0&\kappa_1&0 \\
0&0&0&0 \\
0&0&0&0 \\
\end{array} \right), \;
\sigma_+=\left(\begin{array}{cccc}
0&0&0&0 \\
0&0&0&0 \\
0&\kappa_1&0&0 \\
\kappa_1&0&0&0 \\
\end{array} \right). \nonumber
\end{eqnarray}
Here $\psi(z)=(\psi_1(z),\psi_2(z),\psi_3(z),...,\psi_N(z))^T$ with
$\psi_n(z)=(c_n^{(1)}(z),c_n^{(2)}(z),c_n^{(3)}(z),c_n^{(4)}(z))^T$,
and $c_n^{(j)}(z)$ is the complex field amplitude in the $j$th
waveguide of $n$th quadrimer for $j=1,2,3,4$ and $n=1,2,3,...,N$.
The diagonal blocks $h_n$ describe isolated quadrimers, and the off-diagonal block matrix $\sigma_{\pm}$ describe the coupling between
two nearest quadrimers, where $\gamma$ is the gain/loss parameter, $A$ is the modulation amplitude and $\omega$ is the modulation frequency. The intra-dimer coupling strength is $\nu$, which can be tuned by adjusting the distance
between the waveguides along $y$-axis $\Delta y$. The intra-quadrimer coupling strength $\kappa$
and the inter-quadrimer coupling strength $\kappa_1$ can be tuned by intermittently adjusting the distance
between the waveguides along $x$-axis $\Delta x$ \cite{agrawal}.
Without
loss of generality, we choose $\nu=1$ to set the energy scale.
To simplify, we will set all the parameters are dimensionless throughout this paper.

Under PBC, by implementing a Fourier transform
\begin{eqnarray} \label{Fourier_transform}
c_n^{(1)}&=&\frac{1}{\sqrt{N}}\sum_{q}e^{iq(4n-3)}c_{1,q},  \nonumber \\
c_n^{(2)}&=&\frac{1}{\sqrt{N}}\sum_{q}e^{iq(4n-2)}c_{2,q},  \nonumber \\
c_n^{(3)}&=&\frac{1}{\sqrt{N}}\sum_{q}e^{iq(4n-1)}c_{3,q},  \nonumber \\
c_n^{(4)}&=&\frac{1}{\sqrt{N}}\sum_{q}e^{iq(4n)}c_{4,q}, \nonumber
\end{eqnarray}
one can obtain the Hamiltonian in momentum space
\begin{eqnarray} \label{qHamiltonian1}
H_q(z)=\left[\begin{matrix}
i\gamma & \nu e^{iq} & 0 &\kappa_a&\\
\nu e^{-iq} & -i \gamma & \kappa_b^* & 0 &\\
0 & \kappa_b & -A \sin(\omega z) & \nu e^{iq} &\\
\kappa_a^* & 0 & \nu e^{-iq} & A \sin(\omega z) &
\end{matrix}
\right],
\end{eqnarray}
where $\kappa_a=\kappa e^{i3q}+\kappa_1 e^{-iq}$, $\kappa_b=\kappa e^{-iq}+\kappa_1 e^{i3q}$ and $q$ denotes quasi-momentum.
Our system also can be regarded as
two-coupled non-Hermitian SSH chains with periodical modulation or a periodically modulated non-Hermitian SSH$_{4}$ model with long rang coupling.
Obviously, our system has two types
of periodic character, the transverse periodic structure along $x$-axis and longitudinal periodic modulation along $z$-axis \cite{benisty15}.
Meanwhile, its Hamiltonian is characterized by two types
of the $PT$ symmetry, the local and global ones. We say
that the system is locally $PT$-symmetric if the isolated quadrimer
is $PT$-symmetric in the limit $\kappa_1=0$. Obviously, the Hamiltonian $H_1(z)$ is $PT$-symmetric due to $[H_1(z), PT] = 0$, where $P$ is a space-reversal
linear operator
\begin{eqnarray}
P=\left(\begin{array}{cccc}
0&1&0&0 \\
1&0&0&0 \\
0&0&0&1 \\
0&0&1&0 \\
\end{array} \right), \nonumber
\end{eqnarray}
and time operator $T$ reverses the propagation direction: $T: i\rightarrow -i, z\rightarrow -z$. On the other hand,
we say that the system is globally $PT$-symmetric if the
infinite array (\ref{dyequation}) with the matrix $H_q(z)$ in (\ref{qHamiltonian1}) is $PT$-
symmetric for $\kappa_1\neq0$ \cite{morfon17,konotop12}. The array in Fig. 1 consists of
quadrimers which have unbroken local $PT$ symmetry, at
least for small $\gamma$.

\begin{figure}
\includegraphics[width=\columnwidth]{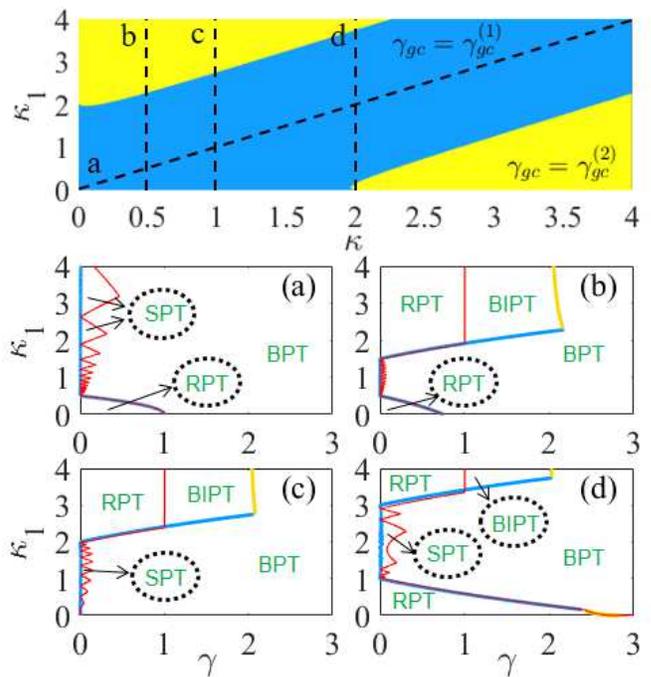}
\caption{
(Color online) (Upper) Selection of the conditions  $\gamma_{gc}^{(1)}$ and $\gamma_{gc}^{(2)}$ in the parameter  plane ($\kappa$, $\kappa_1$). Phase diagram for the system in the parameter plane $(\gamma,\kappa_1)$ for (a) $\kappa=\kappa_1$, (b) $\kappa=0.5$,
(c) $\kappa=1$ and (d) $\kappa=2$. Light blue and yellow lines are analytical results
obtained by the formula (\ref{qcrit}) under PBC
and red lines are numerical results obtained from the Hamiltonian (\ref{model}) under OBC.
The other parameters are chosen as $N=20$, $A=0$ and $\nu$=1.} \label{fig2}
\end{figure}

\begin{figure}
\includegraphics[width=\columnwidth]{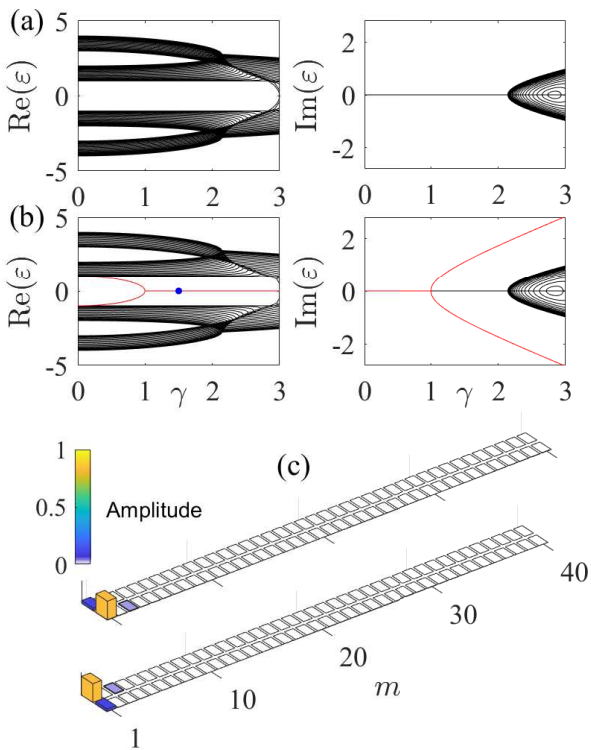}
\caption{(Color online) (a) and (b) Real and imaginary parts of the eigenvalues as a function of $\gamma$ for system under PBC (top) and OBC (bottom).
(c) Representative eigenstates of zero-energy at $\gamma=1.5$ (blue dot in (b)).
The other parameters are $A=0$, $\kappa_1=2.5$, $\kappa=0.5$, $N=20$ and $\nu=1$.} \label{fig4}
\end{figure}

\section{Boundary induced global $PT$-symmetry-breaking}
In this section, we show how the boundary, system size and inter-quadrimer coupling affect the global $PT$ symmetry of the system. In subsection A, we show how to obtain the
phase diagram via analyzing the eigenvalues of system under PBC and OBC.
In subsection B,
we show the energy spectra and dynamics in the three typical phases:
R$PT$, BI$PT$, and B$PT$ phases.

\subsection{Phase diagram}
We use the usual unmodulated system as a reference system.
For $A=0$, the eigenvalues can be given as
\begin{eqnarray} \label{spectequation}
E=\pm\sqrt{D+\nu^2-\frac{\gamma^2}{2}\pm\frac{1}{2}\sqrt{4D(4\nu^2-\gamma^2)+\gamma^4}},
\end{eqnarray}
with $D=\kappa^2+\kappa_{1}^{2}+2\kappa \kappa_1 \cos(4q)$ and $q\in[-\frac{\pi}{4},\frac{\pi}{4}]$. Then the critical value of global $PT$-symmetry-breaking transition is determined by below conditions:
\begin{eqnarray} \label{qcrit}
\gamma_{gc}=\Big \{
\begin{array}{c}
\pm(D-\nu^2)/\nu; \\
\pm\sqrt{2D-2\sqrt{D^2-4D\nu^2}} \\
\end{array}.
\end{eqnarray}
To distinguish, we label $\gamma_{gc}^{(1)}=|(D-\nu^2)/\nu|$ and  $\gamma_{gc}^{(2)}=\Big|\sqrt{2D-2\sqrt{D^2-4D\nu^2}}\Big|$. The selection of the conditions $\gamma_{gc}^{(1)}$ and $\gamma_{gc}^{(2)}$ in the parameter plane ($\kappa$, $\kappa_1$) is given in Fig. (\ref{fig2}).  Since the system is invariant under the transformation $\kappa_1\leftrightarrow\kappa$ under PBC, below
we will consider the case of fixing
$\kappa\leq2$ and increasing $\kappa_1$. The intersection point between two conditions is determined by
\begin{eqnarray} \label{tion}
4\nu^2(\kappa_{1}+\kappa)^2=\frac{[(\kappa_{1}-\kappa)^2-\nu^2]^4}{[(\kappa_{1}-\kappa)^2-\nu^2]^2-4\nu^4}.
\end{eqnarray}
For a fixed $\kappa$, the Eq. (\ref{tion}) may has several real root. To analyze, we define the biggest genuine solution of Eq. (\ref{tion}) as
$\kappa_{1}^{I}$. Therefore for $\kappa\leq2$,
when $\kappa_{1}<\kappa_{1}^{I}$ the critical value of global $PT$-symmetry-breaking transition is determined by condition $\gamma_{gc}^{(1)}$.
When $\kappa_{1}\geq\kappa_{1}^{I}$, the critical value of global $PT$-symmetry-breaking transition is determined by condition $\gamma_{gc}^{(2)}$, as shown in Fig. (\ref{fig2}). It is important to note that
the limited value of $\gamma_{gc}^{(2)}$ always approach to 2 as the increases of $\kappa_{1}$. Therefore,
the maximal critical value of global $PT$-symmetry-breaking transition under PBC can be restored to $\gamma_{gc}\simeq2$ by increasing $\kappa_{1}$. In addition, as shown in Refs. \cite{bendix10}, our unmodulated system under OBC can
support the pairs of degenerate eigenstates that
are the symmetric and anti-symmetric superpositions of two identical well-localized states
centered symmetrically near the opposite boundaries of the system,  and these pairs form
effective dimers, exactly as the single $PT$-symmetric dimer with the critical value
$\gamma_{c}=1$. Therefore,
the maximal critical value of global $PT$-symmetry-breaking transition under OBC only can be restored to $\gamma_{gc}=1$ by increasing $\kappa_{1}$. This implies that global $PT$ symmetry of system under OBC is broken, whereas it still is unbroken under PBC for certain parameter $\kappa_{1}$, and hence the boundary can induce global $PT$-symmetry-breaking. By setting $\gamma_{gc}^{(1)}=1$, the critical point of boundary influencing global $PT$-symmetry is given as
\begin{eqnarray} \label{crittion}
\kappa_{1}^{c}&=&-\kappa\cos(4q) \nonumber \\
&+&\frac{1}{2}\sqrt{-2\kappa^2+4\nu+4\nu^2+2\kappa^2\cos(8q)}.
\end{eqnarray}
Therefore,
the condition of the boundary influencing global $PT$ symmetry is  $\kappa_1\geq\kappa_{1}^{c}$ and $1\leq\gamma\leq2$.
For $\nu=1$,
the condition becomes $\kappa_1\geq\kappa+\sqrt{2}$ and $1\leq\gamma\leq2$.

The above results indicate that for a fixed $\kappa$, depending on values of $\kappa_1$ and $\gamma$, four different situations are  possible: i) robust global-$PT$-symmetry (R$PT$), where global $PT$-symmetries are  unbroken both under PBC and OBC;
ii) boundary influencing global-$PT$-symmetry (BI$PT$), where global $PT$ symmetry is broken under OBC, whereas it still is unbroken under PBC;
iii) system size affecting global-$PT$-symmetry (S$PT$), where global $PT$ symmetry is unbroken under OBC, whereas it is broken under PBC;
iv) broken global-$PT$-symmetry (B$PT$), where global $PT$ symmetries are broken both under PBC and OBC.
We show the phase diagram of global $PT$ symmetry in the parameter plane $(\gamma,\kappa_1)$ for different parameters $\kappa$ in Fig.\ref{fig2} (a)-(d). For the homogeneous coupling case of $\kappa_1=\kappa$,
due to there is not
boundary effect, there does not exist BI$PT$ phase. In addition, because the limited value of $\gamma_{gc}^{(1)}$ and $\gamma_{gc}^{(2)}$ always approach to zero as the increases of $\kappa_{1}$ for $\kappa=\kappa_1$, the global $PT$ symmetry always is destroyed by increasing
$\kappa_1$. Thus, the phase space $(\gamma,\kappa_1)$ can be divided into three domains consisting of R$PT$,  S$PT$ and B$PT$ as shown in Fig.\ref{fig2} (a). For inhomogeneous coupling case of $\kappa\neq\kappa_1$, the phase space $(\gamma,\kappa_1)$ can be divided into multiple domains consisting of R$PT$, BI$PT$, S$PT$ and B$PT$ as shown in Figs.\ref{fig2} (b)-(d).
A important feature of the phase diagram is the existence of the triple point that correspond to values $(\gamma=1,\kappa_1=\kappa+\sqrt{2})$, where three phases of R$PT$, BI$PT$ and B$PT$ touch. Thus depending on how parameters $\gamma$ and $\kappa_1$ change in vicinity of the triple point, breaking of global $PT$ symmetry can occur in two different ways: R$PT$$\rightarrow$ B$PT$, or R$PT$$\rightarrow$BI$PT$$\rightarrow$B$PT$. Depending on
the critical value of spontaneous local $PT$-symmetry-breaking transition $\gamma_{lc}$, which is relation to the ration of $\pm(\nu^2-\kappa^2)/\nu$,
the domain of R$PT$ phase under parameter
region of $\kappa_1<\kappa$ and $\gamma<\gamma_{lc}$ shrinks with the increase of $\kappa$ for $\kappa<\nu$ and enlarges with the increase of $\kappa$ for $\kappa>\nu$ [Figs.\ref{fig2} (b) and \ref{fig2}(d)]. In particular, it disappears at $\kappa=\nu$ [Fig.\ref{fig2} (c)].

\subsection{The energy spectra and dynamics in the different phases}
To see clearly how the spectra change in different phases with the increase of $\gamma$, we show the real and imaginary parts of the eigenvalues as a function of $\gamma$ for the system under PBC and OBC. A typical example is displayed in Figs. \ref{fig4} by choosing        $\kappa=0.5$ and $\kappa_1=2.5$, which can undergo two phase transitions from R$PT$ phase to BI$PT$ phase and from BI$PT$ phase to  B$PT$ phase as $\gamma$ continuously increase. It is clear that
in the R$PT$ phase the system has a purely real spectrum when $\gamma<1$, however under OBC, a pair of isolated edge states with real spectrum begin to emerge when $\gamma\geq0.5$. After undergoing a phase transition from R$PT$ phase to BI$PT$ phase at $\gamma=1$,
the real parts of the eigenvalues of this pair of edge states twofold degenerate zero-energy level, meanwhile its imaginary parts split into one pair of conjugated imaginary values with nonzero.  Therefore, in the BI$PT$ phase,
there exists a pair of isolated zero-energy edge states with purely imaginary energy
eigenvalues localized at the left boundary, whereas other $4N-2$ eigenvalues of bulk states are real, see Fig.  \ref{fig4}(b). Therefore, the essence of the first phase transition from R$PT$ phase to BI$PT$ phase is the transition of a pair of edge states of the system from real eigenvalues to complex eigenvalues.
For convenience, we define this critical value as $\gamma_{gc}^{e}$.
This is why the boundary can break global $PT$ symmetry in our system. Of course, a pair of isolated edge states with purely real  energy
eigenvalues localized at the right boundary also can occur with the increase of $\kappa_1$, but they are not zero-energy. After undergoing second phase transition from BI$PT$ phase to B$PT$ phase at $\gamma\simeq2$, the real parts of $4N-2$ ($4N$ under PBC) eigenvalues of bulk states begin to degenerate,
meanwhile their imaginary parts begin to split into many pair of conjugated imaginary values with nonzero. Therefore, the essence of the second phase transition from BI$PT$ phase to B$PT$ phase is the transition of bulk states of the system from real eigenvalues to complex eigenvalues. Similarly, we define this critical value as $\gamma_{gc}^{b}$.

\begin{figure}
\includegraphics[width=\columnwidth]{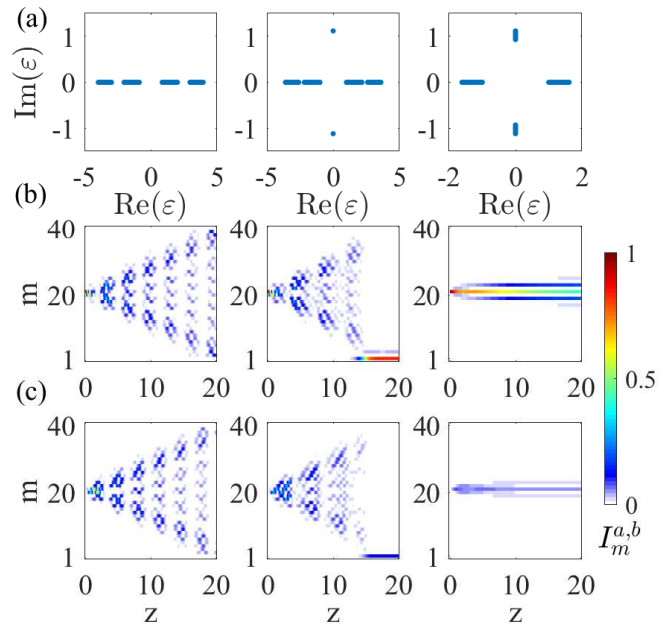}
\caption{(Color online) (Upper) Eigenvalues distribution in the
complex-energy plane. The wave packet evolution
in above (middle) and below (bottom) row waveguides, respectively. In the left
column we choose parameters in R$PT$ phase, such as $\kappa=0.5$, $\kappa_1=2.5$ and
$\gamma=0.5$. In the middle column we choose parameters in BI$PT$ phase, such as $\kappa=0.5$, $\kappa_1=2.5$ and
$\gamma=1.5$. In the
right column we choose parameters in B$PT$ phase, such as  $\kappa=0.5$, $\kappa_1=0.5$ and
$\gamma=1.5$. The other parameters are chosen as $N=20$ and
$\nu$=1.} \label{fig5}
\end{figure}

Through numerical integration, we analyze the light
propagations of the
coupled-mode system (\ref{dyequation}) in three different phases with $N=20$. The light propagation sensitively
depends upon the eigenvalues. Stationary light
propagations of bounded intensity oscillations appear if
all eigenvalues are real. Local light propagations
of unbounded intensity oscillations appear if at least one of
the eigenvalues is complex. To do this, we firstly define the light intensities in above row waveguides $I_m^{a}=|c_n^{j}(z)|^2$ for $m=1, 2, 3, ..., 2N$, $n=1, 2, 3, ..., N$ and $j=$ 1, and 4, where $I_m^{a}=|c_n^{1}(z)|^2$ for $m$
equaling odd, and $I_m^{a}=|c_n^{4}(z)|^2$ for $m$
equaling even. Similarly, the light intensities in below row waveguides are defined as
$I_m^{b}=|c_n^{j}(z)|^2$ for $j=$ 2, and 3, where $I_m^{b}=|c_n^{2}(z)|^2$ for $m$
equaling odd, and $I_m^{b}=|c_n^{3}(z)|^2$ for $m$
equaling even. In Fig. \ref{fig5}, we give the light propagation using
single-site excitation of away from the edge waveguide. If we choose parameters in R$PT$ phase, the light always diffracts irrespectively
of its input cell,
see the left
column in Fig. \ref{fig5}. If we
choose parameters in BI$PT$ phase, local light propagating along the edge route appear at the left edge of system,
see the middle
column in Fig. \ref{fig5}. The local
light propagating along the left edge route is a direct signature
of the BI$PT$ phase. This implies that our system can be used to realize single edge-mode lasing or robust one-way edge mode transport \cite{parto18}.
If we
choose parameters in B$PT$ phase, the light propagating always localize at its input cell, and propagate
with unbounded intensity oscillations,
see the right
column in Fig. \ref{fig5}. Therefore, our numerical
simulations of the coupled-mode system (\ref{dyequation})
perfectly confirm the BI$PT$ phase predicted by our analytical results.

\begin{figure}
\includegraphics[width=\columnwidth]{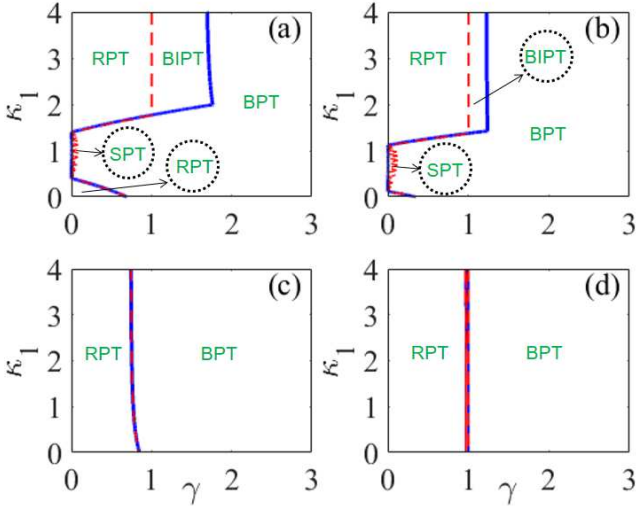}
\caption{
(Color online) Phase diagram of the system in the parameter plane $(\gamma,\kappa_1)$ for different modulation parameters (a) $A/\omega=0.6$, (b) $A/\omega=1$,
(c) $A/\omega=1.5$ and (d) $A/\omega=2.4$. Blue lines are numerical results obtained from the Hamiltonian (\ref{eqHamiltonian1}), and red lines are numerical results obtained from the Hamiltonian (\ref{model}).
The other parameters are chosen as $N=20$, $\kappa=0.5$, $\omega=10$ and $\nu$=1.} \label{fig6}
\end{figure}


\begin{figure}
\includegraphics[width=\columnwidth]{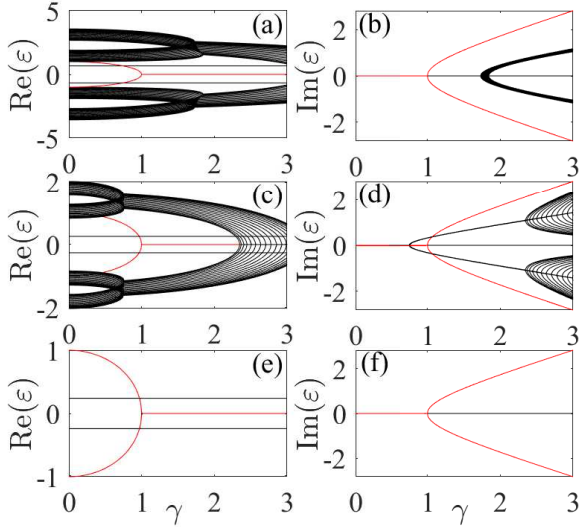}
\caption{(Color online) Real and imaginary parts of the quasienergies of the system under OBC as a function of $\gamma$ for $A/\omega=0.6$ (top), 1.5 (middle) and 2.4 (bottom).
The other parameters are $\omega=10$, $\kappa_1=2.5$, $\kappa=0.5$, $N=20$ and $\nu=1$.} \label{fig7a}
\end{figure}

\begin{figure}
\includegraphics[width=\columnwidth]{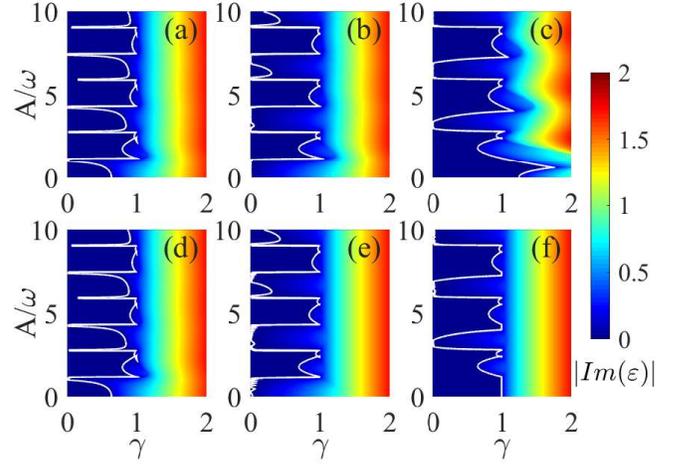}
\caption{
(Color online) The imaginary parts of the quasienergies $|Im(\varepsilon)|$ as a function of $A/\omega$ and $\gamma$ under PBC (top) and OBC (bottom)
with different coupling strengthes $\kappa_1=0.1$ (left), $\kappa_1=1$ (middle) and $\kappa_1=2$ (right). The other parameters are chosen as $N=20$, $\kappa=0.5$, $\omega=10$ and
$\nu$=1.} \label{fig7}
\end{figure}

\section{Manipulation of global $PT$ symmetry}
In this section, we will explore how
to manipulate the global $PT$ symmetry via the periodic modulation.
According to the Floquet theorem, similar to the Bloch states, the Floquet
states of the modulated system (\ref{dyequation})
satisfy $\{\psi_1(z),\psi_2(z),\psi_3(z),...,\psi_n(z)\}=e^{-i \epsilon z}\{\widetilde{\psi}_1(z),\widetilde{\psi}_2(z),\widetilde{\psi}_3(z),...,\widetilde{\psi}_n(z)\}$. Here, the propagation constant $\epsilon$ is
called the quasienergy, and the complex amplitudes $\widetilde{\psi}_n(z)$ are periodic with the modulation period $T=2\pi/\omega$.
Then the quasienergies and eigenfunctions are given by
$\mathcal{F}\widetilde{\psi}_n=\epsilon \widetilde{\psi}_n$,
with the Floquet operator
$\mathcal{F}=-i\frac{d}{dz}+H_{N}(z)$.
Under high-frequency Floquet analysis, the correspondingly effective Hamiltonian can be written as
\begin{eqnarray} \label{eqHamiltonian1}
H_{q}^{eff}=\left[\begin{matrix}
i\gamma & \nu e^{iq} & 0 &\kappa_a \tilde{J}&\\
\nu e^{-iq} & -i \gamma & \kappa_b^*\tilde{J}^* & 0 &\\
0 & \kappa_b \tilde{J} & 0 & \nu e^{iq} \tilde{Q} &\\
\kappa_a^* \tilde{J}^* & 0 & \nu e^{-iq} \tilde{Q}^* & 0 &
\end{matrix}
\right],
\end{eqnarray}
with
\begin{eqnarray} \label{6}
\tilde{J}&=&\sum_{k=-\infty}^{\infty}(i)^k J_k(A/\omega) e^{ ik\omega z}, \nonumber\\
\tilde{Q}&=&\sum_{k=-\infty}^{\infty}(i)^k J_k(2A/\omega) e^{ ik\omega z}. \nonumber
\end{eqnarray}
The modulus of $\tilde{J}$ and $\tilde{Q}$ depend on the values of $A/\omega$, and can change between zero and one. In particular, the modulus $|\tilde{J}|$ equals
zero at some specific values of $A/\omega$ (such as $A/\omega \simeq$ $2.4$ and $5.52$), and the modulus $|\tilde{Q}|$ equals zero at $A/\omega \simeq$ $1.2$ and $2.76$.
Then the spectra $\varepsilon$ are given by the roots
of the secular equation
\begin{eqnarray} \label{secularequation}
&&\varepsilon^4 +[\gamma^2 - \nu^2(1+ \tilde{Q}^2) - \tilde{J}^2(|\kappa_a|^2+|\kappa_b|^2)] \varepsilon^2 \nonumber \\
&&+i\gamma \tilde{J}^2(|\kappa_b|^2-|\kappa_a|^2)\varepsilon-\tilde{J}^2\nu^2 \tilde{Q} e^{-i2q}(\kappa_a\kappa_b+\kappa_a^*\kappa_b^*)    \nonumber \\
&&+(\nu^4-\gamma^2\nu^2)\tilde{Q}^2+ \tilde{J}^4|\kappa_a|^2|\kappa_b|^2=0.
\end{eqnarray}
From Eq. (\ref{secularequation}), we can see
for a fixed $\gamma$, $\kappa$ and $\kappa_1$,
the spontaneous global $PT$-symmetry-breaking transition can be manipulated by tuning modulation parameter, because the modulus
$|\tilde{J}|$ and $|\tilde{Q}|$ periodically change from zero to one along with
the modulation parameter $A/\omega$.
For the special value $\tilde{J}=0$, the quasienergies become as $\varepsilon=\pm \tilde{Q} \nu$ and $\pm\sqrt{\nu^2-\gamma^2}$, and the critical value of global $PT$-symmetry-breaking transition is $\gamma_{gc}=\nu$. In this situation,
the system decouples $N$ uncoupled $PT$-symmetric dimers, and hence the critical value of $PT$-symmetry-breaking transition is determined by single $PT$-symmetric dimer.
For the the special value $\tilde{Q}=0$, the quasienergies become as
\begin{eqnarray} \label{qspectequation}
\varepsilon=\pm\sqrt{\frac{2D\tilde{J}^2+\Gamma\pm\sqrt{\Gamma(4D\tilde{J}^2+\Gamma)}}{2}}, \nonumber
\end{eqnarray}
with $\Gamma=\nu^2-\gamma^2$. The critical value of global $PT$-symmetry-breaking transition is $\gamma_{gc}=\nu$. Therefore, for the special values $\tilde{J}=0$ or $\tilde{Q}=0$,
the critical value of global $PT$-symmetry-breaking transition $\gamma_{gc}=\nu=1$ is independent of parameters $\kappa$, $\kappa_1$ and boundary condition. Then in the parametric region $\kappa_1>\kappa+\sqrt{2}$ and $1<\gamma\leq2$, the critical value $\gamma_{gc}^{b}$ is adjusted from 2 to 1, as $A/\omega$ increases from 0 to 1.2. Furthermore, the critical value $\gamma_{gc}^{b}$ can be adjusted to less than 1 for $1.2< A/\omega<2.4$.
This implies that the domain of the BI$PT$ phase can be manipulated narrow and even disappeared by tuned $A/\omega$.

To verify above analytical results,
we show the phase diagram of global $PT$ symmetry in the parameter plane $(\gamma,\kappa_1)$ for different modulation parameters $A/\omega$ in Fig.\ref{fig6}. As an example, we choose the unmodulated case in Fig.\ref{fig2} (b) as a reference system. Obviously, the presence of periodic modulation
modifies previous physical picture. Firstly, it clearly shows
that the domain of the BI$PT$ phase can be manipulated narrow and even disappeared by tuning $A/\omega$, and hence the triple point also vanish.
Second,
the domain of B$PT$ phase under parameter
region of $\kappa_1<\kappa+\sqrt{2}$ and $\gamma<1$ can be adjusted R$PT$ phase with the increase of $A/\omega$.       Then
the phase space $(\gamma,\kappa_1)$ only can be divided into two domains consisting of R$PT$ and B$PT$ for $1.2\leq A/\omega\leq2.4$, where the critical value $\gamma_{gc}$ not only is independent of parameters $\kappa$ and $\kappa_1$, but also independent of boundary condition, as shown in Figs.\ref{fig6}(c) and (d).
To see clearly how the spectra change in different phases with the increase of $\gamma$ for different modulation parameters $A/\omega$, we show the real and imaginary parts of the quasienergies as a function of $\gamma$ for the system under OBC. A typical example is displayed in Figs. \ref{fig7a} by choosing modulation parameters $A/\omega=$ 0.6, 1.5 and 2.4. It is clearly show that the critical value $\gamma_{gc}^{e}$ is independent of modulation parameter $A/\omega$, but the critical value $\gamma_{gc}^{b}$ can be adjusted small by tuning $A/\omega$. The critical value $\gamma_{gc}^{b}$ change from 2 to 1 as $A/\omega$ increases from 0 to 1.2, and $\gamma_{gc}^{b}$ can be adjusted to less than 1 for $1.2<A/\omega<2.4$. Therefore, when  $\gamma_{gc}^{b}\leq\gamma_{gc}^{e}$ for $1.2\leq A/\omega\leq2.4$, a pair of isolated zero-energy edge states
embed into bulk states, and then the BI$PT$ phase disappears.
This is why BI$PT$ phase
can be manipulated by tuning modulation parameter in our system.

To show the parameter dependence of global $PT$ symmetry from another angle, we show the quasienergies
$|Im(\varepsilon)|$ as a function of $A/\omega$ and $\gamma$ for different coupling parameters ($\kappa$, $\kappa_1$) both for PBC and OBC. As an example, we choose three sets of coupling parameters that initially be in R$PT$ phase ($\kappa=0.5$, $\kappa_1=0.1$), B$PT$ phase ($\kappa=0.5$, $\kappa_1=1$) and BI$PT$ phase ($\kappa=0.5$, $\kappa_1=2$), as shown in  Fig.\ref{fig7}. It is clearly see
that whether or not the system has initial global $PT$ symmetry, the periodic modulation not only can restore the broken global $PT$ symmetry, but also can control it by tuning modulation amplitude.
Differently, the global $PT$-symmetry only can be adjusted to the critical value $\gamma_{gc}=1$ for parametric region that initially be in R$PT$ phase, B$PT$ phase and BI$PT$ phase of the system under OBC, whereas it can be adjusted to $\gamma_{gc}>1$ for parametric region that initially be in BI$PT$ phase of the system under PBC.  Therefore, for our modulated system of fixed $\gamma$, it is possible to observe
the spontaneous global $PT$-symmetry-breaking transition by
tuning $A/\omega$. It is important to note that the global property of transverse periodic structure of such a 2D optical array can be manipulated
by only tuning modulation amplitude of longitudinal periodic modulation.

\section{Conclusion and discussion}
In summary, we have investigated the static and dynamical property of global $PT$ symmetry in 2D arrays of periodically modulated $PT$-symmetric quadrimer waveguides.
For unmodulated case with inhomogeneous inter- and intra- quadrimer coupling strength $\kappa_1\neq\kappa$, in addition to conventional global $PT$-symmetric phases and $PT$-symmetry-breaking phase, we find
that there is exotic phase where global $PT$ symmetry is broken under OBC, whereas it still is $PT$-symmetric under PBC.
The boundary of phase is analytically given as $\kappa_1\geq\kappa+\sqrt{2}$ and $1\leq\gamma\leq2$, where there exists a pair of zero-energy edge states with purely imaginary energy
eigenvalues localized at the left boundary, whereas other $4N-2$ eigenvalues are real.
The parametric dependence of the spontaneous global  $PT$-symmetry-breaking is analytically and numerically
explored for the modulated array. Because critical value $\gamma_{gc}^{e}$ is independent of modulation parameter $A/\omega$, and the critical value $\gamma_{gc}^{b}$ can be adjusted small by tuning $A/\omega$,
the domain of the BI$PT$ phase can be manipulated narrow and even disappeared by tuning $A/\omega$. More interestingly,
whether or not the array has initial global $PT$ symmetry, periodic modulation not only can restore the broken global $PT$ symmetry, but also can control it by tuning modulation amplitude. Therefore, the global property of transverse periodic structure of a 2D optical array can be manipulated
by only tuning modulation amplitude of longitudinal periodic modulation.
Our results
provide a promising approach for designing and manipulating optical material and may have specific technological importance.

With currently available techniques, it is possible
to realize our model and observe our theoretical predictions
with experiments. Our proposed structure can be demonstrated experimentally
in numerous optical systems \cite{ruter, bersch, weiman2017}. For instance, in
photonics, one can use the femtosecond direct writing method
\cite{sza2010} to realize a 2D array of $PT$-symmetric photonic
coupled waveguide.
Periodic modulations can be introduced by harmonic modulations
of the real refractive index or periodic curvatures
along the propagation direction \cite{long2009, gara2012,zeuner2015}.
For such
a 2D optical array with periodic modulation,
spontaneous global $PT$-symmetry-breaking transition may be observed by adjusting the modulation
parameter. In addition, it is also possible
to apply our model and method for designing some optical devices, such as single edge-mode laser and robust one-way edge mode transport.

\begin{acknowledgments}
H. Zhong is thankful to Chaohong Lee for enlightening suggestions and helpful discussions. This work is supported by the National Natural Science Foundation of China under Grant No. 11805283 and No. 11725417, the Hunan Provincial Natural Science Foundation under Grants No. 2019JJ30044, and the Talent project of Central South University of Forestry and Technology under Grant No. 2017YJ035.
\end{acknowledgments}

\end{document}